\documentclass[a4paper,oneside]{article}
\usepackage{graphicx}
\usepackage{amsmath}
\usepackage{babel}

\newtheorem{theorem}{Theorem}

\newtheorem{remark}[theorem]{Remark}

\input{tcilatex}

\begin{document}

\author{\textit{S. Manoff}\thanks{%
Department of Theoretical Physics, Institute for Nuclear Research and
Nuclear Energy, Blvd. Tzarigradsko Chaussee 72, 1784 - Sofia, Bulgaria}}
\title{Auto-parallel equation as Euler-Lagrange's equation over spaces with affine
connections and metrics}
\date{\textit{e-mail address: smanov@inrne.bas.bg}}
\maketitle

\begin{abstract}
The auto-parallel equation over spaces with affine connections and metrics [$%
(\overline{L}_n,g)$-spaces] is considered as a result of the application of
the method of Lagrangians with covariant derivatives (MLCD) on a given
Lagrangian density.
\end{abstract}

\section{\strut Introduction}

In (pseudo) Riemannian spaces without torsion ($V_n$-spaces) the geodesic
equation (identical with the auto-parallel equation $\nabla _uu=0$) \cite
{Schmutzer} can be found on the ground of the variation of an action $S$
identified with the parameter $s$ of a curve interpreted as its length $s$%
\begin{equation}
S=\int ds+s_0\text{ }:\delta S=0\Rightarrow \nabla _uu=0\text{ , with }%
\nabla =\Gamma =\{\,\,\,\}\text{ ,}  \label{0.1}
\end{equation}

\noindent where $\Gamma =\{\,\,\}$ is the symmetric Levi-Civita affine
connection and 
\begin{eqnarray}
\nabla _{u}u &=&u^{i}\,_{;j}\cdot u^{j}\cdot \partial _{i}\text{ ,\thinspace
\thinspace \thinspace \thinspace \thinspace \thinspace \thinspace \thinspace
\thinspace \thinspace }u^{i}\,_{;j}=u^{i}\,_{,j}+\Gamma _{kj}^{i}\cdot u^{k}%
\text{ ,\thinspace \thinspace \thinspace \thinspace \thinspace }
\label{0.1a} \\
\text{\thinspace \thinspace \thinspace }u &=&u^{i}\cdot \partial
_{i}=u^{\alpha }\cdot e_{\alpha }\in T(M)\text{ ,\thinspace \thinspace } \\
\text{\thinspace \thinspace \thinspace }\Gamma _{kj}^{i} &=&\{_{kj}^{i}\}=%
\frac{1}{2}\cdot g^{im}\cdot (g_{jk,m}+g_{km,j}-g_{kj,m})\text{ , }
\label{0.2} \\
\text{\thinspace \thinspace \thinspace }g_{ij} &=&g_{ji}\text{ , \thinspace
\thinspace }g_{ij;k}=0\text{ , \ \ \  }i\text{, }j\text{, }...=1,2,...,n%
\text{ ,} \\
\text{ \ \ \  }dimM &=&n\text{ \thinspace \thinspace ,\thinspace \thinspace
\thinspace \thinspace \thinspace \thinspace \thinspace \thinspace \thinspace
\thinspace }u^{i}\,_{,j}=\partial u^{i}/\partial x^{j}\text{ ,\thinspace
\thinspace \thinspace \thinspace \thinspace \thinspace \thinspace }%
u^{i}=dx^{i}/ds\text{ ,}  \notag
\end{eqnarray}

\begin{eqnarray}
&&\{\partial _{i}\}\text{ is a co-ordinate (holonomic) contravariant basis
in }T(M)\text{, } \\
&&\{e_{\alpha }\}\text{ is a non-co-ordinate (non-holonomic) contravariant
basis in }T(M)\text{ .}  \notag
\end{eqnarray}

The same method has been used for finding out the geodesic equation in a
space with contravariant and covariant affine connections (whose components
differ not only by sign) and metrics \cite{Manoff-1} [called $(\overline{L}%
_n,g)$-space] \cite{Manoff-2}. Since the geodesic equation (interpreted as
an equation for motion of a moving free spinless test particle in an
external gravitational field) differs from the auto-parallel equation in a $(%
\overline{L}_n,g)$-space or in a space with an affine connection and metrics
[$(L_n,g)$-space], the old question arises as what is the right equation for
description of a free moving particle in a $(\overline{L}_n,g)$ - or a $%
(L_n,g)$ -space: the geodesic equation (G) or the auto-parallel equation
(A). The most authors \textit{believe} that the geodesic equation is the
more appropriate equation. One of their major arguments is that the geodesic
equation is related to a variational principle (as a basic principle in
classical physics) in contrast to the auto-parallel equation.

At first, the so called A-G problem has induced discussions in the case of
(pseudo) Riemannian spaces with torsion ($U_n$-spaces, Riemann-Cartan
spaces) used for describing the gravitational interaction in the
Einstein-Cartan-etc. theories of gravitation and in the theory of defects in
crystals \cite{Kleinert-1}. The classical action (\ref{0.1}) has been
considered by Kleinert and collaborators \cite{Kleinert-2} - \cite
{Kleinert-7}, as well as by other authors \cite{Saa-1}-\cite{Saa-3} \cite
{Fiziev-1}, \cite{Fiziev-2} in a co-ordinate basis and in a non-co-ordinate
basis in a $V_n$-space leading in the first basis to the geodesic equation
in a $V_n$-space in a co-ordinate basis and in the second basis leading to
the geodesic equation in a non-co-ordinate basis, where the structure
coefficients $C_{\alpha \beta }\,^\gamma $ \{$[e_\alpha ,e_\beta ]=C_{\alpha
\beta }\,^\gamma \cdot e_\gamma $\} are related to the torsion tensor $%
T_{\alpha \beta }\,^\gamma =\Gamma _{\beta \alpha }\,^\gamma -\Gamma
_{\alpha \beta }\,^\gamma -C_{\alpha \beta }\,^\gamma $ \cite{Sa} and the
equation is interpreted as an auto-parallel equation. Analogous
consideration by the use of the embedding of a Riemannian space with torsion
in an Euclidean space (without torsion) \cite{Kleinert-8} has been made.
Both the methods have been critically evaluated and the main disadvantages
of the proposed variational principle have been summarized by S\'{a} \cite
{Sa}.

Recently, it has been proved by Iliev \cite{Iliev-1}-\cite{Iliev-7} and
latter by Hartley \cite{Hartley} that every given affine (linear) connection
in a $(L_n,g)$-space could vanish at a point or on a trajectory in a
properly chosen co-ordinate or non-co-ordinate basis in this space. If the
space has torsion, then the basis should be chosen as a non-co-ordinate
basis. That means that in a $(L_n,g)$-space the auto-parallel equation could
be written in the forms 
\begin{equation}
\nabla _uu=u^i\,_{;j}\cdot u^j\cdot \partial _i=u^\alpha \,_{/\beta }\cdot
u^\beta \cdot e_\alpha =0\text{ ,}  \label{0.2a}
\end{equation}
\begin{equation}
u^i\,_{;j}\cdot u^j=u^i\,_{,j}\cdot u^j+\Gamma ^i\,_{kj}\cdot u^k\cdot u^j=0%
\text{ ,}  \label{2b}
\end{equation}
\begin{equation}
\frac{d^2x^i}{ds^2}+\Gamma _{kj}^i\cdot \frac{dx^k}{ds}\cdot \frac{dx^j}{ds}%
=0\text{ \thinspace \thinspace \thinspace \thinspace \thinspace for
\thinspace \thinspace \thinspace \thinspace \thinspace }\Gamma _{kj}^i\neq 0%
\text{ \thinspace \thinspace ,}  \label{0.2c}
\end{equation}
\begin{equation}
u^\alpha \,_{/\beta }\cdot u^\beta =e_\beta u^\alpha +\Gamma _{\gamma \beta
}^\alpha \cdot u^\gamma \cdot u^\beta =0\text{ ,}  \label{0.2d}
\end{equation}
\begin{equation}
\frac{d^2\overline{x}^\alpha }{ds^2}=0\text{ \thinspace \thinspace
for\thinspace \thinspace \thinspace \thinspace \thinspace \thinspace
\thinspace }\Gamma _{\beta \gamma }^\alpha =0\text{ \thinspace \thinspace
\thinspace \thinspace \thinspace and\thinspace \thinspace \thinspace }%
T_{\alpha \beta }\,^\gamma \neq 0\text{ ,}  \label{0.2e}
\end{equation}

\noindent where $u^\alpha =$ $d\overline{x}^\alpha /ds=A_j\,^\alpha \cdot
u^j $ are the components of the tangent vector $u=u^\alpha \cdot e_\alpha $
to the trajectory $x^i(\tau )$ in a non-co-ordinate (non-holonomic)
contravariant basis $\{e_\alpha \}\in T(M)$ and\thinspace $\overline{x}%
^a=A_i\,^\alpha \cdot dx^i$.

Therefore, in every $(L_n,g)$-space the auto-parallel equation (\ref{0.2a})
could be written in a special basis as an equation for motion of a moving
free spinless test particle (\ref{0.2e}). ''This fact leads to the
conclusion that the equivalence principle in Einstein's theory of
gravitation (ETG) can be considered only as a physical interpretation of a
corollary of the mathematical apparatus and its validity can be extended to
all spaces with affine connections and metrics. Even if a differentiable
manifold has two connections (whose components differ not only by sign) for
tangent and cotangent vector fields, the principle of equivalence is
fulfilled for one of the two types of vector fields. Therefore, every
differentiable manifold with affine connections and metrics can be used as a
model for space-time in which the equivalence principle holds. On this
ground, a free moving spinless test particle in a suitable basic system in a 
$(L_n,g)$- or $(\overline{L}_n,g)$-space $(n=4)$ will fulfill an equation
identical with that for the motion of a free moving spinless test particle
in Newtonian mechanics or in special relativity \cite{Manoff-2}.'' The
auto-parallel equation (interpreted in ETG, the special relativity theory,
and in Newtonian mechanics as an equation for the motion of a free moving
spinless test particle, and identical with the geodesic equation) can be
considered as a generalization of the equation of a free moving spinless
test particle in the case of $(L_n,g)$- or $(\overline{L}_n,g)$-spaces. The
question that arises for the case of these type of spaces is the same as for
the case of $U_n$-spaces: ''Is it possible the auto-parallel equation to be
derived from a variational principle?'' This is an important question,
inducing an important task, because there are evidences that $(L_n,g)$- and $%
(\overline{L}_n,g)$-spaces can have similar structures as the $V_n$-spaces
for describing dynamical systems and the gravitational interaction. In such
type of spaces one could use Fermi-Walker transports \cite{Manoff-3}, \cite
{Manoff-4} conformal transports \cite{Manoff-5}, \cite{Manoff-6} and
different frames of reference \cite{Manoff-7}. All these notions appear as
generalizations of the corresponding notions in $V_n$-spaces.

On this ground, in our opinion, the failure in finding an appropriate
Lagrangian formalism for obtaining the auto-parallel equation is related to
the attempts of using analogous variational expressions for the action $S$
as in the case of the geodesic equation. In $(\overline{L}_n,g)$- and $%
(L_n,g)$-spaces the auto-parallel equation has much more complicated
structure (related to torsion and nonmetricity) than the geodesic equation 
\cite{Manoff-2}. The auto-parallel equation $\nabla _uu=0$ should be
fulfilled for a given affine connection independent of a given metric. It
should not depend on a metric in a manifold $M$ (in contrast to the geodesic
equation in $V_n$-spaces). An affine connection determines transports of
other contravariant vector fields $\{\xi \in T(M)\}$ along an auto-parallel
vector field $u$. The fact that $u$ is an auto-parallel vector field should
in some way influence the transport of a set of vector fields $\{\xi \}$
along $u$ (if $u$ is parallel transported, then $\xi $ is transported along $%
u$ in determined way). On the other side, an auto-parallel contravariant
vector field $u$ induces (on the ground of an existing metric in a
differentiable manifold $M$) additional structures such as the orthogonal to
it sub space $T^{\perp u}(M)$ which could also be taken into account if we
wish to find the auto-parallel equation on the basis of a variational
principle. In a variational principle for obtaining the auto-parallel
equation $\nabla _uu=0$ the mentioned above circumstances should be taken
into account. If we use the method of Lagrangians with covariant derivatives
(MLCD) we could find a solution of the G-A problem.

The method of Lagrangians with covariant derivatives is a field theoretical
method, worked out as a Lagrangian formalism for tensor field theories over $%
(L_n,g)$- and $(\overline{L}_n,g)$-spaces \cite{Manoff-8}, \cite{Manoff-9}.\
In this sense, it is more general than the Lagrangian formalism in classical
mechanics or in general relativity \cite{Manoff-10}.

In this paper we consider a possible representation of the auto-parallel
equation by the use of a Lagrangian formalism based on the method of
Lagrangians with covariant derivatives. The method is entirely different
from the methods used by Kleinert et all which are applicable in $U_n$%
-spaces but not very useful in $(L_n,g)$- or $(\overline{L}_n,g)$-spaces,
where the nonmetricity could play as important role as the torsion. In
Section 2 we recall some well known facts about the canonical
representations of the parallel and the auto-parallel equations over spaces
with affine connections and metrics. In Section 3 we consider degenerated
Lagrangian densities and their corresponding Euler-Lagrange's equations. In
Section 4 the auto-parallel equation is obtained and investigated on the
basis of the MLCD as an Euler-Lagrange's equation related to a degenerated
Lagrangian density with respect to a preliminary given contravariant vector
field $\xi $. The last Section 5 comprises some concluding remarks. The most
considerations are given in details (even in full details) for those readers
who are not familiar with the investigated problems.

\section{Canonical representation of the parallel and the auto-parallel
equations}

\subsection{Canonical representation of a parallel equation}

Let us now recall some well known facts from the differential geometry of
manifolds \cite{Bishop}. Let a congruence $x^i(\tau ,\lambda )$ be given
described by the two parameters $\tau $ and $\lambda $ and by the tangent
vector fields $\xi $, $\eta \in T(M)$: 
\begin{equation}
u:=\frac \partial {\partial \tau }=\frac{\partial x^i}{\partial \tau }\cdot
\partial _i\text{ , \thinspace \thinspace \thinspace \thinspace \thinspace
\thinspace \thinspace \thinspace \thinspace \thinspace and \thinspace
\thinspace \thinspace \thinspace \thinspace \thinspace \thinspace \thinspace
\thinspace }\xi :=\frac \partial {\partial \lambda }=\frac{\partial x^j}{%
\partial \lambda }\cdot \partial _j\text{ }  \label{0.3}
\end{equation}

\noindent respectively. Let us consider a parallel transport of the vector
field $\xi $ along the vector field $u$%
\begin{equation}
\nabla _u\xi =f\cdot \xi \text{ , \thinspace \thinspace \thinspace
\thinspace \thinspace \thinspace \thinspace \thinspace }f\in C^r(M)\text{ .}
\label{3.1}
\end{equation}

An equation of this type is called recurrent equation (or recurrent relation
for the vector field $\xi $) \cite{Sinjukov}. Three types of invariance of
this equation could be found.

(a) Invariance (of type $A$) with respect to a change of the co-ordinates
(charts) or invariance (of type $B$) with respect to a change of the bases
in the manifold $M$. These types ($A$ and $B$) of invariance are obvious
because they follow from the index-free form of the equation.

(b) Form invariance with respect to a change of the vector $\xi $ with a
collinear to it vector $\eta :=\varphi \cdot \xi $ [$\varphi =\varphi
(x^k)\neq 0$].

Proof. If we express the vector field $\xi $ with its equivalent form $\xi
=\varphi ^{-1}\cdot \eta $ in the above equation, we will get 
\begin{equation}
\nabla _u(\varphi ^{-1}\cdot \eta )=u(\varphi ^{-1})\cdot \eta +\varphi
^{-1}\cdot \nabla _u\eta =f\cdot \varphi ^{-1}\cdot \eta \text{ .}
\label{0.4}
\end{equation}

After the necessary transformation of the single terms, the equation for $%
\eta $ can be written in the form 
\begin{equation}
\nabla _{u}\eta =\overline{f}\cdot \eta \text{ , \thinspace \thinspace
\thinspace \thinspace \thinspace \thinspace \thinspace }\overline{f}%
:=f-u(\log \varphi ^{-1})\text{ \ .}  \label{0.5}
\end{equation}

Therefore, the parallel equation for $\xi $ does not change its form by the
change of $\xi $ with a collinear vector field $\eta $.

(c) Form invariance with respect to a change of the parameter $\lambda $
determining $\xi $.

Proof. The change of the parameter $\lambda $ with a new parameter $\sigma
=\sigma (\lambda )$ with $\lambda =\lambda (\sigma )$, and $\frac{d\sigma }{%
d\tau }=0$, leads to the relations 
\begin{equation}
\xi ^i=\frac{\partial x^i}{\partial \lambda }=\frac{\partial x^i}{\partial
\sigma }\cdot \frac{d\sigma }{d\lambda }=\overline{\xi }\,^i\cdot \frac{%
d\sigma }{d\lambda }\text{ , \thinspace \thinspace \thinspace \thinspace
\thinspace \thinspace \thinspace \thinspace \thinspace \thinspace \thinspace
\thinspace \thinspace \thinspace }\overline{\xi }\,^i=\frac{\partial x^i}{%
\partial \sigma }\text{ ,\thinspace \thinspace \thinspace \thinspace
\thinspace \thinspace \thinspace \thinspace \thinspace \thinspace \thinspace 
}\frac{d\sigma }{d\lambda }:\neq 0\text{ .}  \label{0.6}
\end{equation}
\begin{equation}
\xi ^i\,_{,j}=\left( \frac{\partial x^i}{\partial \lambda }\right)
_{,j}=\left( \frac{\partial x^i}{\partial \sigma }\cdot \frac{d\sigma }{%
d\lambda }\right) _{,j}=\left( \frac{\partial x^i}{\partial \sigma }\right)
_{,j}\cdot \frac{d\sigma }{d\lambda }+\frac{\partial x^i}{\partial \sigma }%
\cdot \left( \frac{d\sigma }{d\lambda }\right) _{,j}\text{ ,}  \label{0.7}
\end{equation}
\begin{equation}
\xi ^i\,_{,j}\cdot u^j=\xi ^i\,_{,j}\cdot \frac{\partial x^j}{\partial \tau }%
=\frac{\partial \overline{\xi }\,^i}{\partial \tau }\frac{d\sigma }{d\lambda 
}+\overline{\xi }\,^i\cdot \frac \partial {\partial \tau }\left( \frac{%
d\sigma }{d\lambda }\right) \text{ .}  \label{0.8}
\end{equation}

Since $\sigma $ and $\lambda $ are independent of the parameter $\tau $ the
second term at the right of the equations vanishes. Therefore, 
\begin{equation}
\xi ^i\,_{,j}\cdot u^j=\xi ^i\,_{,j}\cdot \frac{\partial x^j}{\partial \tau }%
=\frac{\partial \overline{\xi }\,^i}{\partial \tau }\frac{d\sigma }{d\lambda 
}=\frac{\partial \xi ^i}{\partial \tau }\text{ }  \label{0.9}
\end{equation}

\noindent and the parallel equation for $\xi $ will be expressed in terms of 
$\overline{\xi }$ in the form 
\begin{equation}
\xi ^i\,_{;j}\cdot u^j=\xi ^i\,_{,j}\cdot u^j+\Gamma _{jk}^i\cdot \xi
^j\cdot u^k=(\frac{\partial \overline{\xi }\,^i}{\partial \tau }+\Gamma
_{jk}^i\cdot \overline{\xi }\,^j\cdot u^k)\cdot \frac{d\sigma }{d\lambda }%
=f\cdot \overline{\xi }\,^i\cdot \frac{d\sigma }{d\lambda }\text{ }.
\label{0.10}
\end{equation}

Therefore, we have 
\begin{equation}
\frac{\partial \overline{\xi }\,^i}{\partial \tau }+\Gamma _{jk}^i\cdot 
\overline{\xi }\,^j\cdot u^k=\overline{\xi }\,^i\,_{;j}\cdot u^j=f\cdot 
\overline{\xi }\,^i\text{ \thinspace \thinspace \thinspace \thinspace
\thinspace }\cong \,\,\,\,\,\,\,\,\,\nabla _u\overline{\xi }=f\cdot 
\overline{\xi }\text{ , \thinspace \thinspace \thinspace \thinspace
\thinspace }\overline{\xi }\,=\overline{\xi }\,^i\cdot \partial _i\text{ .}
\label{0.11}
\end{equation}

\subsection{Canonical representation of an auto-parallel equation}

Let us now consider the auto-parallel equation $\nabla _uu=f\cdot u$ as a
special case of a parallel equation for $\xi =u$, $f=f(x^k(\tau ))$. In this
case 
\begin{equation}
\lambda =\tau \text{ \thinspace \thinspace \thinspace ,\thinspace \thinspace
\thinspace \thinspace }u=\frac d{d\tau }\text{ \thinspace \thinspace
\thinspace \thinspace , \thinspace \thinspace \thinspace \thinspace }u^i=%
\frac{dx^i}{d\tau }\text{ \thinspace \thinspace \thinspace and \thinspace
\thinspace \thinspace \thinspace \thinspace }\sigma =\sigma (\tau )\text{ ,
\thinspace \thinspace \thinspace }\tau =\tau (\sigma )\text{ , }\frac{%
d\sigma }{d\tau }\neq 0\text{ .}  \label{0.12}
\end{equation}

Then 
\begin{equation}
u^i=\frac{dx^i}{d\sigma }\cdot \frac{d\sigma }{d\tau }=\overline{u}^i\cdot 
\frac{d\sigma }{d\tau }\text{ , \thinspace \thinspace \thinspace \thinspace
\thinspace }\overline{u}^i=\frac{dx^i}{d\sigma }\text{ ,}  \label{0.13}
\end{equation}
\begin{equation}
u^i\,_{,j}\cdot u^j=\frac{d\overline{u}^i}{d\sigma }\cdot \left( \frac{%
d\sigma }{d\tau }\right) ^2+\overline{u}^i\cdot \frac{d^2\sigma }{d\tau ^2}%
\text{ ,}  \label{0.14}
\end{equation}
\begin{equation}
u^i\,_{;j}\cdot u^j-f\cdot u^i=\left( \frac{d\sigma }{d\tau }\right) ^2\cdot
\left( \frac{d\overline{u}^i}{d\sigma }+\Gamma _{jk}^i\cdot \overline{u}%
^j\cdot \overline{u}^k\right) +\overline{u}^i\cdot \left( \frac{d^2\sigma }{%
d\tau ^2}-f(\tau )\cdot \frac{d\sigma }{d\tau }\right) =0\text{ .}
\label{0.15}
\end{equation}

One can chose as condition for determining the function $\sigma =\sigma
(\tau )$ as a function of $\tau $ the vanishing of the last term of the
above equation 
\begin{equation}
\frac{d^2\sigma }{d\tau ^2}-f(\tau )\cdot \frac{d\sigma }{d\tau }=0\text{ .}
\label{0.16}
\end{equation}

The last equation is of the type 
\begin{equation}
y^{\prime }-f\cdot y=0\text{ , \thinspace \thinspace where\thinspace
\thinspace }\,\,\,\,\,\,\,\,\,y=\frac{d\sigma }{d\tau }\text{ ,}%
\,\,\,\,\,\,\,\,\,\,y^{\prime }=\frac{d^2\sigma .}{d\tau ^2}\text{
.\thinspace \thinspace }  \label{0.17}
\end{equation}

Then

\begin{equation}
\sigma =\sigma _0+\sigma _1\cdot \int \exp \left( \int f(\tau )\cdot d\tau
\right) \cdot d\tau \text{ , \thinspace \thinspace \thinspace \thinspace
\thinspace }\sigma _0=\text{ const.,\thinspace \thinspace \thinspace
\thinspace \thinspace \thinspace \thinspace \thinspace }\sigma _1=\text{
const.}  \label{0.18}
\end{equation}

After the introduction of the new parameter $\sigma =\sigma (\tau )$ (called
canonical parameter), the auto-parallel equation will have the form 
\begin{equation}
\nabla _{\overline{u}}\overline{u}=0\text{ \thinspace \thinspace \thinspace }%
\cong \,\,\,\,\overline{u}^i\,_{;j}\cdot \overline{u}^j=0\,\,\,\,\,\text{%
\thinspace , \thinspace \thinspace \thinspace \thinspace \thinspace
\thinspace \thinspace \thinspace \thinspace }\overline{u}=\frac d{d\sigma }%
\text{ ,\thinspace \thinspace \thinspace \thinspace \thinspace \thinspace
\thinspace \thinspace \thinspace \thinspace }\overline{u}^i=\frac{dx^i}{%
d\sigma }\text{ .}  \label{0.19}
\end{equation}

The last equation for $\overline{u}$ is called auto-parallel equation in
canonical form.

\section{\strut Degenerated Lagrangian densities and Euler-Lagrange's
equations}

A degenerated Lagrangian density with respect to field variables $V^A\,_B$
is a Lagrangian density $\mathbf{L}$ of the type 
\begin{equation}
\mathbf{L}=\sqrt{-d_g}\cdot L(K^A\,_B\text{, }K^A\,_{B;i}\text{, }%
K^A\,_{B;i;j}\text{, }V^C\,_D)\text{ ,}  \label{Ch 20 3.1}
\end{equation}

\noindent where 
\begin{equation}
V^C\,_D\neq K^A\,_B\text{ .}  \label{0.20}
\end{equation}

The functions $K^A\,_B(x^k)$ and $V^C\,_D(x^k)$ are components of tensor
fields with finite rank. The symbol $_{;k}$ is denoted the covariant
derivative, defined by the use of given affine connections, with respect to
a co-ordinate $x^k$ or to a basis $e_k$ ($\partial _k$) . Therefore, a
degenerated Lagrangian invariant $L$ is an invariant depending only on the
field variables $V^C\,_D$ and not depending on their first and second
(partial or covariant) derivatives, i.e. 
\begin{equation}
L=L(K^A\,_B\text{, }K^A\,_{B;i}\text{, }K^A\,_{B;i;j}\text{, }V^C\,_D)\text{
.}  \label{Ch 20 3.2}
\end{equation}

If we consider the field variables $V^C\,_D$ as dynamic field variables,
then the functional variation of the Lagrangian density $\mathbf{L}$,
leading to the corresponding Euler-Lagrange's equations, has the form \cite
{Manoff-9} 
\begin{equation}
\frac{\delta \mathbf{L}}{\delta V^C\,_D}=\sqrt{-d_g}\cdot \frac{\partial L}{%
\partial V^C\,_D}\,\,\,\,\,\,\,\,\,\,\,\,\,\,\text{for\thinspace \thinspace
\thinspace \thinspace }\,\,\,V^C\,_D\neq g_{ij}\,\,(\neq K^A\,_B),
\label{Ch 20 3.3}
\end{equation}

\begin{equation}
\frac{\delta \mathbf{L}}{\delta g_{kl}}=\sqrt{-d_g}\cdot (\frac{\partial L}{%
\partial g_{kl}}+\frac 12\cdot L\cdot g^{\overline{k}\overline{l}})\text{
\thinspace \thinspace \thinspace \thinspace for \thinspace \thinspace
\thinspace \thinspace }V^C\,_D=g_{kl}\text{ \thinspace \thinspace }(\neq
K^A\,_B)\text{ .}  \label{Ch 20 3.4}
\end{equation}

The Euler-Lagrange equations (if they could exist) for $V^C\,_D$ could be
found in the form

(a) For $V^C\,_D\neq g_{kl}$: 
\begin{equation}
\frac{\partial L}{\partial V^C\,_D}=0\text{ .}  \label{Ch 20 3.5}
\end{equation}

(b) For $V^C\,_D=g_{kl}$: 
\begin{equation}
\frac{\partial L}{\partial g_{kl}}+\frac 12\cdot L\cdot g^{\overline{k}%
\overline{l}}=0\text{ .}  \label{Ch 20 3.6}
\end{equation}

Let us now consider both the cases separately to each other.

\subsection{Degenerated Lagrangian densities with respect to non-metric
field variables}

Let $V^C\,_D\neq g_{kl}$ be given fulfilling the Euler-Lagrange equations 
\begin{equation}
\frac{\partial L}{\partial V^C\,_D}=G\,_C\,^D=0\text{ .}  \label{0.21}
\end{equation}
One can distinguish sub cases following as solutions of the equations (\ref
{Ch 20 3.5}):

(a) $L$ is independent of the field variables $V^C\,_D$. This conclusion
from the equations contradicts to the prerequisite for the structure of $L$.

(b) $L$ is linearly dependent on $V^C\,_D$. Then $L$ could be written in the
form 
\begin{eqnarray}
L &=&L_0+F(K^A\,_B\text{, }K^A\,_{B;i}\text{, }K^A\,_{B;i;j})\cdot
G_C\,^D\,(K^A\,_B\text{, }K^A\,_{B;i}\text{, }K^A\,_{B;i;j})\cdot V^C\,_D%
\text{ ,}  \label{Ch 20 3.7} \\
L_0 &=&\text{ const.}  \notag
\end{eqnarray}

The Euler-Lagrange equations for $V^C\,_D$ degenerate to conditions for the
other field variables $K^A\,_B$%
\begin{equation}
\frac{\partial L}{\partial V^C\,_D}=F\cdot G_C\,^D=0\text{ .}
\label{Ch 20 3.8}
\end{equation}

The last equations, among with the Euler-Lagrange equations for the field
variables $K^A\,_B$, form a system of differential equations for all field
variables in $L$ considered as dynamic field variables. Here $V^C\,_D$ take
the role of Lagrangian multipliers. If $K^A\,_B$ are not considered as
dynamic field variables, i.e. if they are assumed as preliminary given
non-dynamic field variables, then the equations (\ref{Ch 20 3.8}) appear as
additional conditions (constrains) for $K^A\,_B$. If (\ref{Ch 20 3.8}) are
fulfilled, then $L=L_0=$ const. Furthermore, if we interpret $L$ as the
pressure $p=L$ in a physical system \cite{Manoff-11}, \cite{Manoff-12} then
the existence of Euler-Lagrange equations for $V^C\,_D$ leads to
establishing of a constant or vanishing pressure [$L=L_0=p_0=$ const. $(\neq
0$, $=0)$] in the system. Therefore, if we wish to consider a system with $%
p=p_0=$ const.$(\neq 0$, $=0)$, we can introduce a Lagrangian invariant $L$
of type (b) and then we can find the Euler-Lagrange equations for the
dynamic field variables $V^C\,_D$ and their corresponding energy-momentum
tensors by the use of the method of Lagrangians with covariant derivatives
(MLCD).

(c) $L$ does not depend linearly on $V^C\,_D$. Then $L$ could have the
general form 
\begin{equation}
L=L(K^A\,_B\text{, }K^A\,_{B;i}\text{, }K^A\,_{B;i;j}\text{, }V^C\,_D)\text{
.}  \label{0.22}
\end{equation}

The Euler-Lagrange equations for $V^C\,_D$ degenerate in this case to
algebraic equations for $V^C\,_D$%
\begin{equation}
G_C\,^D=\frac{\partial L}{\partial V^C\,_D}=0\text{ .}  \label{Ch 20 3.9}
\end{equation}

\subsection{Degenerated Lagrangian densities with respect to metric field
variables}

Let $V^C\,_D=g_{kl}$ be given fulfilling the Euler-Lagrange equations 
\begin{equation}
\frac{\partial L}{\partial g_{kl}}+\frac 12\cdot L\cdot g^{\overline{k}%
\overline{l}}=0\text{ .}  \label{0.23}
\end{equation}

The invariant $L$ has to obey the condition 
\begin{equation}
\frac{\partial L}{\partial g_{ij}}\cdot g_{ij}+\frac n2\cdot L=0\text{ ,}
\label{Ch 20 3.10}
\end{equation}

i.e. it should be a homogeneous function of degree $h=-\frac n2$ with
respect to the metric field components $g_{ij}$%
\begin{equation}
\frac{\partial L}{\partial g_{ij}}\cdot g_{ij}=-\frac n2\cdot L\text{ .}
\label{Ch 20 3.11a}
\end{equation}

We can distinguish some sub cases induced by the possible solutions of the
Euler-Lagrange equations.

(a) $L$ is linearly dependent on $g_{ij}$, i.e. $L$ could be represented in
the form 
\begin{equation}
L=L_0+k\cdot F\cdot G^{\overline{i}\overline{j}}\cdot g_{ij}\text{ ,}
\label{Ch 20 3.11}
\end{equation}

where 
\begin{equation}
L_0=\text{ const.,\thinspace \thinspace \thinspace \thinspace \thinspace
\thinspace \thinspace \thinspace }k=\text{ const.,\thinspace \thinspace
\thinspace \thinspace \thinspace \thinspace \thinspace \thinspace \thinspace
\thinspace \thinspace }G^{\overline{i}\overline{j}}=G^{\overline{j}\overline{%
i}}\text{ .}  \label{0.24}
\end{equation}

From (\ref{Ch 20 3.11}), it follows that 
\begin{equation}
\frac{\partial L}{\partial g_{ij}}=k\cdot F\cdot G^{\overline{i}\overline{j}%
}\,  \label{Ch 20 3.12}
\end{equation}

and 
\begin{equation}
\frac{\partial L}{\partial g_{ij}}\cdot g_{ij}=k\cdot F\cdot G^{\overline{i}%
\overline{j}}\cdot g_{ij}=L-L_0=-\frac n2\cdot
L:\,\,\,\,\,\,\,\,\,\,\,\,\,(1+\frac n2)\cdot L=L_0\text{ .}
\label{Ch 20 3.13}
\end{equation}

If $L_0=0$, then either $n=-2$ (which contradicts to the condition $%
dimM=n\geq 1$), or $L=0$. Then, 
\begin{equation}
\frac{\partial L}{\partial g_{ij}}\cdot
g_{ij}=0:\,\,\,\,\,\,\,\,\,\,\,\,\,\,\,\,\,\,\,\,\,\,\,k\cdot F\cdot G^{%
\overline{i}\overline{j}}=0\text{ .}  \label{Ch 20 3.14}
\end{equation}

If $L_0\neq 0$, then 
\begin{equation}
L=\frac 2{n+2}\cdot L_0=\text{ const.,\thinspace \thinspace \thinspace
\thinspace \thinspace \thinspace \thinspace \thinspace \thinspace \thinspace
\thinspace \thinspace \thinspace \thinspace }k\cdot F\cdot G^{\overline{i}%
\overline{j}}\text{\thinspace \thinspace \thinspace \thinspace }+\frac
1{n+2}\cdot L_0\cdot g^{\overline{i}\overline{j}}=0\text{ , \thinspace
\thinspace \thinspace }  \label{Ch 20 3.15}
\end{equation}
\begin{equation}
g^{ij}=-\,\,\,(n+2)\cdot \frac k{L_0}\cdot F\cdot G^{ij}\text{ .}
\label{Ch 20 3.16}
\end{equation}

The explicit form of $g_{ij}$ is determined by the functions $F$ and $G^{ij}$
as functions of the field variables $K^A\,_B$ and their first and second
covariant derivatives.

(b) $L$ does not dependent linearly on $g_{ij}$. At the same time, $L$
should fulfil the homogeneous conditions (\ref{Ch 20 3.11a}). The
Euler-Lagrange equations for $g_{ij}$ appear as algebraic equations for $%
g_{ij}$. Together with the Euler-Lagrange equations for the field variables $%
K^A\,_B$ they determine a system of differential equations for the field
variables $(K^A\,_B$, $g_{ij})$.

\section{Auto-parallel equation as Euler-Lagrange's equation}

\subsection{Euler-Lagrange's equation and auto-parallel equation for a
vector field $u$}

Every Lagrangian invariant $L$ could be interpreted as the pressure $p$ of
the considered physical system in a $4$-dimensional space with affine
connections and metrics used as a model of space-time \cite{Manoff-8}, \cite
{Manoff-11}.

Let us define a Lagrangian invariant in the form 
\begin{equation}
L:=p=p_0+a_0\cdot \rho \cdot e+h_0\cdot g(\nabla _u\rho u,\xi )\text{ ,}
\label{3.11}
\end{equation}

\noindent where 
\begin{eqnarray}
p_0 &=&\text{ const., \thinspace \thinspace \thinspace \thinspace \thinspace 
}a_0=\text{ const., \thinspace \thinspace \thinspace \thinspace }h_0=\text{
const., \thinspace \thinspace \thinspace }  \label{0.25a} \\
\rho &=&\rho (x^k)\in C^r\text{ , \thinspace \thinspace \thinspace
\thinspace \thinspace }r\geq 2\text{ , \thinspace \thinspace \thinspace
\thinspace \thinspace \thinspace \thinspace \thinspace \thinspace \thinspace 
}u\text{, }\xi \,\in T(M)\text{ ,}  \label{0.25b} \\
e &=&g(u,u)\neq 0\text{ , \thinspace \thinspace \thinspace \thinspace
\thinspace \thinspace \thinspace }g=g_{ij}\cdot dx^i.dx^j\text{ .}
\label{0.25}
\end{eqnarray}

The pressure $p$ could also be written in the form 
\begin{equation}
p=p_0+f\cdot \rho +b\cdot (u\rho )\text{ ,}  \label{3.12}
\end{equation}

\noindent where 
\begin{eqnarray}
f &=&a_0\cdot e+h_0\cdot g(a,\xi )\text{ , \thinspace \thinspace \thinspace
\thinspace \thinspace \thinspace \thinspace \thinspace \thinspace \thinspace
\thinspace \thinspace \thinspace \thinspace \thinspace }a=\nabla _uu\text{
\thinspace \thinspace \thinspace \thinspace ,}  \label{0.26a} \\
b &=&h_0\cdot l\text{ , \thinspace \thinspace \thinspace \thinspace
\thinspace \thinspace \thinspace \thinspace \thinspace \thinspace \thinspace 
}l=g(u,\xi )\text{ .}  \label{0.26}
\end{eqnarray}

The constant $p_0$ is a constant pressure, $\rho \cdot e$ is interpreted as
the kinetic energy density, $u$ is the velocity of the points in the system, 
$\xi $ is a contravariant vector transported along $u$.

In a co-ordinate basis $f$ and $b$ will have the forms respectively 
\begin{eqnarray}
f &=&g_{\overline{k}\overline{l}}\cdot (a_0\cdot u^k\cdot u^l+h_0\cdot u^k%
\text{ }_{;m}.u^m\cdot \xi ^l)\text{ ,}  \label{3.13a} \\
b &=&h_0\cdot g_{\overline{k}\overline{l}}\cdot u^k\cdot \xi ^l\text{ .}
\label{3.13}
\end{eqnarray}

By the use of the method of Lagrangians with covariant derivatives we can
find the Euler-Lagrange equations for the field variables $\rho $, $u$, $\xi 
$, and $g$ as well as the corresponding energy-momentum tensors. Moreover,
we can consider every sub set of the set $\{\rho $, $u$, $\xi $, $g\}$ as
dynamic field variables (for which the Euler-Lagrange's equations should be
found), where the rest of the field variables are considered as non-dynamic
field variables. In the further investigations we will consider all field
variables $\{\rho $, $u$, $\xi $, $g\}$ as dynamic field variables. At that,
the Lagrangian invariant $L$ appears as a degenerated Lagrangian invariant
with respect to the metric tensor components $g_{ij}$ and the components $%
\xi ^i$ of the contravariant vector $\xi $.

For finding out the Euler-Lagrange equations for $\rho $, $u$, $\xi $, and $%
g $, we have to find the explicit form of some auxiliary expressions. Let us
make a list of them. 
\begin{equation}
\frac{\partial e}{\partial u^i}=2\cdot g_{\overline{i}\overline{k}}\cdot
u^k=2\cdot u_{\overline{i}}\text{ ,\thinspace \thinspace \thinspace
\thinspace \thinspace \thinspace \thinspace \thinspace \thinspace \thinspace
\thinspace \thinspace \thinspace \thinspace \thinspace \thinspace \thinspace
\thinspace \thinspace \thinspace \thinspace \thinspace }\frac{\partial e}{%
\partial g_{ij}}=u^{\overline{i}}\cdot u^{\overline{j}}\text{ ,}
\label{0.27}
\end{equation}
\begin{equation}
\frac{\partial g(a,\xi )}{\partial u^i}=g_{\overline{k}\overline{l}}\cdot
\xi ^l\cdot u^k\,_{;i}\text{ , \thinspace \thinspace \thinspace \thinspace
\thinspace \thinspace \thinspace \thinspace \thinspace \thinspace \thinspace
\thinspace \thinspace \thinspace \thinspace \thinspace \thinspace \thinspace
\thinspace \thinspace }\frac{\partial g(a,\xi )}{\partial u^i\,_{;j}}=g_{%
\overline{i}\overline{k}}\cdot \xi ^k\cdot u^j\text{ ,\thinspace \thinspace
\thinspace \thinspace \thinspace \thinspace \thinspace \thinspace \thinspace
\thinspace \thinspace \thinspace \thinspace \thinspace \thinspace }\frac{%
\partial g(a,\xi )}{\partial \xi ^i}=g_{\overline{i}\overline{k}}\cdot .a^k%
\text{ ,}  \label{0.28}
\end{equation}
\begin{equation}
\frac{\partial g(a,\xi )}{\partial g_{ij}}=\frac 12\cdot (a^{\overline{i}%
}\cdot \xi ^{\overline{j}}+a^{\overline{j}}\cdot \xi ^{\overline{i}})\text{ ,%
}  \label{0.29}
\end{equation}
\begin{equation}
\frac{\partial b}{\partial \rho }=0\text{ , \thinspace \thinspace \thinspace
\thinspace \thinspace }\frac{\partial f}{\partial \rho }=0\text{ ,
\thinspace \thinspace \thinspace \thinspace \thinspace }\frac{\partial b}{%
\partial \rho _{,i}}=\frac{\partial b}{\partial (\rho _{,i})}=0\text{ ,
\thinspace }  \label{0.30}
\end{equation}
\begin{equation}
\frac{\partial b}{\partial u^i}=h_0\cdot g_{\overline{i}\overline{k}}\cdot
\xi ^k\text{ ,\thinspace \thinspace \thinspace \thinspace \thinspace
\thinspace \thinspace \thinspace \thinspace }\frac{\partial b}{\partial \xi
^i}=h_0\cdot g_{\overline{i}\overline{k}}\cdot u^k\text{ , \thinspace
\thinspace \thinspace \thinspace \thinspace \thinspace \thinspace \thinspace
\thinspace \thinspace }\frac{\partial b}{\partial g_{ij}}=\frac 12\cdot
h_0\cdot (u^{\overline{i}}\cdot \xi ^{\overline{j}}+u^{\overline{j}}\cdot
\xi ^{\overline{i}})\text{ ,}  \label{0.31}
\end{equation}
\begin{equation}
\frac{\partial (u\rho )}{\partial u^i}=b\cdot \rho _{,i}\text{ , \thinspace
\thinspace \thinspace \thinspace \thinspace \thinspace \thinspace \thinspace
\thinspace \thinspace \thinspace \thinspace \thinspace \thinspace }\delta
u=u^i\,_{;i}\text{ \thinspace \thinspace \thinspace \thinspace \thinspace
\thinspace . \thinspace \thinspace \thinspace \thinspace \thinspace
\thinspace \thinspace \thinspace \thinspace }  \label{0.32}
\end{equation}

\subsubsection{Euler-Lagrange's equation for the scalar function $\protect%
\rho $}

The Euler-Lagrange equation for $\rho $ has the form \cite{Manoff-8} 
\begin{equation}
\frac{\partial p}{\partial \rho }-(\frac{\partial p}{\partial \rho _{,i}}%
)_{;i}+q_i\cdot \frac{\partial \rho }{\partial \rho _{,i}}=0\text{ ,}
\label{3.14}
\end{equation}

\noindent where 
\begin{equation}
\frac{\partial p}{\partial \rho }=f\text{ ,\thinspace \thinspace \thinspace
\thinspace \thinspace \thinspace \thinspace \thinspace \thinspace \thinspace
\thinspace \thinspace \thinspace }\frac{\partial p}{\partial \rho _{,i}}%
=b\cdot u^i\text{ , \thinspace \thinspace \thinspace \thinspace \thinspace
\thinspace \thinspace \thinspace \thinspace \thinspace \thinspace \thinspace
\thinspace }(\frac{\partial p}{\partial \rho _{,i}})_{;i}=ub+b\cdot \delta
u=b_{,i}\cdot u^i+b\cdot u_{\,\,\,\,;i}^i\text{ \thinspace \thinspace
\thinspace \thinspace \thinspace ,}  \label{3.15}
\end{equation}
\begin{equation}
q_i=T_{ki}\,^i-\frac 12\cdot g^{\overline{k}\overline{l}}\cdot
g_{kl;i}+g_k^l\cdot g_{l;i}^k\text{ , \thinspace \thinspace \thinspace
\thinspace \thinspace \thinspace \thinspace \thinspace }q=q_j\cdot u^j\text{
, \thinspace \thinspace \thinspace \thinspace \thinspace \thinspace }%
q_i\cdot \frac{\partial \rho }{\partial \rho _{,i}}=b\cdot q\text{\thinspace
\thinspace \thinspace \thinspace \thinspace \thinspace ,}  \label{3.16}
\end{equation}

After substituting the explicit forms of the single terms in the
Euler-Lagrange's equation, we obtain the explicit form of the
Euler-Lagrange's equation for the scalar function $\rho $ in the form 
\begin{equation}
ub=f+(q-\delta u)\cdot b\text{ \thinspace \thinspace \thinspace ,}
\label{3.17}
\end{equation}

\noindent or in the forms 
\begin{equation}
h_0\cdot ul=a_0\cdot e+h_0\cdot g(a,\xi )+h_0\cdot (q-\delta u)\cdot l\text{
,}  \label{0.33}
\end{equation}
\begin{equation}
l_{,i}\cdot u^i=g_{\overline{k}\overline{l}}\cdot \left\{ \frac{a_0}{h_0}%
\cdot u^k\cdot u^l+[u^k\,_{;m}\cdot u^m\cdot \xi ^l+(q_i\cdot
u^i-u^i\,_{;i})\cdot u^k\cdot \xi ^l]\right\} \text{ .}  \label{3.18}
\end{equation}

It follows from the explicit form of the Euler-Lagrange equation for $\rho $
that this equation does not depend on $\rho $ and could be considered as a
condition for $l=g(u,\xi )$.

\subsubsection{Euler-Lagrange's equations for the contravariant vector field 
$u$}

The Euler-Lagrange equations for the vector field $u$ have the form 
\begin{equation}
\frac{\partial p}{\partial u^i}-(\frac{\partial p}{\partial u^i\,_{;j}}%
)_{;j}+q_j\cdot \frac{\partial p}{\partial u^i\,_{;j}}=0\text{ \thinspace
\thinspace \thinspace ,}  \label{3.19}
\end{equation}

\noindent where 
\begin{equation}
\frac{\partial p}{\partial u^i}=h_0\cdot l\cdot \rho _{,i}+g_{\overline{i}%
\overline{k}}\cdot (2\cdot a_0\cdot \rho \cdot u^k+h_0\cdot \rho _{,j}\cdot
u^j\cdot \xi ^k)+h_0\cdot \rho \cdot g_{\overline{k}\overline{l}}\cdot
u^k\,_{;i}\cdot \xi ^l\text{ ,}  \label{3.20}
\end{equation}
\begin{equation}
\frac{\partial p}{\partial u^i\,_{;j}}=\rho \cdot \frac{\partial f}{\partial
u^i\,_{;j}}=h_0\cdot \rho \cdot g_{\overline{i}\overline{k}}\cdot \xi
^k\cdot u^j\text{ \thinspace \thinspace \thinspace ,}  \label{3.21}
\end{equation}
\begin{equation}
q_j\cdot \frac{\partial p}{\partial u^i\,_{;j}}=h_0\cdot \rho \cdot q\cdot
g_{\overline{i}\overline{k}}\cdot \xi ^k\text{ \thinspace \thinspace
\thinspace ,}  \label{3.22}
\end{equation}
\begin{eqnarray}
(\frac{\partial p}{\partial u^i\,_{;j}})_{;j} &=&h_0\cdot \{g_{\overline{i}%
\overline{k}}\cdot [\rho _{,j}\cdot u^j\cdot \xi ^k+\rho \cdot (\xi
^k\,_{;j}\cdot u^j+u^m\,_{;m}\cdot \xi ^k-g_{l;j}^k\cdot u^j\cdot \xi ^l)]+ 
\notag \\
&&+\rho \cdot (g_{\overline{i}\overline{k}})_{;j}\cdot u^j\cdot \xi ^k\}%
\text{ .}  \label{3.23}
\end{eqnarray}

After substituting the above expressions in (\ref{3.19}), we obtain the
explicit form of the Euler-Lagrange equations for $u^i$ as equations for the
components $\xi ^i$ of the contravariant vector field $\xi $%
\begin{eqnarray}
\xi ^i\,_{;j}\cdot u^j &=&l\cdot (log\,\rho )_{,j}\cdot g^{ji}+2\cdot \frac{%
a_0}{h_0}\cdot u^i+(q-\delta u)\cdot \xi ^i+g_{\overline{k}\overline{l}%
}\cdot u^k\,_{;j}\cdot g^{ji}\cdot \xi ^l-  \notag \\
&&-g^{ij}\cdot (g_{\overline{j}\overline{k}})_{;m}\cdot u^m\cdot \xi
^k+g_{k;j}^i\cdot u^j\cdot \xi ^k\text{ .}  \label{3.24}
\end{eqnarray}

The last equations determine the transport of the contravariant vector field 
$\xi $ along the vector field $u$.

\subsubsection{Euler-Lagrange's equations for the contravariant vector field 
$\protect\xi $}

Since $p$ depends only on the components $\xi ^i$ of the vector field $\xi $
and does not depend on its covariant derivatives, $p$ appears as a
degenerated Lagrangian invariant with respect to the field variables $\xi ^i$%
. The Euler-Lagrange equations for the vector field $\xi $ have the form 
\begin{equation}
\frac{\partial p}{\partial \xi ^i}=0\text{ ,}  \label{3.25}
\end{equation}

where 
\begin{equation}
\frac{\partial p}{\partial \xi ^i}=\rho \cdot \frac{\partial f}{\partial \xi
^i}+(u\rho )\cdot \frac{\partial b}{\partial \xi ^i}\text{ ,\thinspace
\thinspace \thinspace \thinspace \thinspace \thinspace \thinspace \thinspace
\thinspace \thinspace \thinspace \thinspace \thinspace }u\rho =\rho
_{,j}\cdot u^j\text{ ,}  \label{3.26}
\end{equation}
\begin{equation}
\frac{\partial b}{\partial \xi ^i}=h_0\cdot g_{\overline{i}\overline{k}%
}\cdot u^k\text{ \thinspace , \thinspace \thinspace \thinspace \thinspace
\thinspace \thinspace \thinspace \thinspace \thinspace \thinspace \thinspace
\thinspace \thinspace }\frac{\partial f}{\partial \xi ^i}=h_0\cdot g_{%
\overline{i}\overline{k}}\cdot a^k\text{ ,\thinspace \thinspace \thinspace
\thinspace \thinspace }  \label{3.27}
\end{equation}
\begin{equation}
\frac{\partial p}{\partial \xi ^i}=h_0\cdot g_{\overline{i}\overline{k}%
}\cdot [\rho \cdot a^k+(u\rho )\cdot u^k]\text{ .}  \label{3.28}
\end{equation}

The explicit form of the Euler-Lagrange equations for $\xi ^i$ follows in
the form of equations for the components $u^i$ of the contravariant vector
field $u$%
\begin{equation}
a^i=u^i\,_{;j}\cdot u^j=-[(log\,\rho )_{,j}\cdot u^j]\cdot u^i\text{ ,}
\label{3.29}
\end{equation}

\noindent or in index-free form 
\begin{equation}
a=\nabla _uu=f\cdot u\text{ ,\thinspace \thinspace \thinspace \thinspace
\thinspace \thinspace \thinspace \thinspace \thinspace \thinspace \thinspace
\thinspace \thinspace \thinspace \thinspace \thinspace \thinspace \thinspace 
}f=-[(log\,\rho )_{,j}\cdot u^j]\text{ .}  \label{3.30}
\end{equation}

The last equation is exactly the auto-parallel equation for the vector field 
$u$ in a non-canonical form. After changing the parameter $\tau $ of the
curve $x^k(\tau )$, to which $u=\frac d{d\tau }$ is a tangent vector, the
equation could be written in its canonical form $\nabla _{\overline{u}}%
\overline{u}=0$, where 
\begin{equation}
\overline{u}=\frac d{d\sigma }\text{ ,\thinspace \thinspace \thinspace
\thinspace \thinspace \thinspace \thinspace \thinspace \thinspace \thinspace
\thinspace \thinspace }\sigma =\sigma _1\cdot \int \exp \left( \int f(\tau
)\cdot d\tau \right) \cdot d\tau +\sigma _0\text{ , \thinspace \thinspace
\thinspace \thinspace \thinspace \thinspace \thinspace \thinspace \thinspace
\thinspace }\sigma _1=\text{ const., }\sigma _0=\text{ const.,}  \label{3.31}
\end{equation}
\begin{equation}
\sigma =\rho _0\cdot \int \frac 1\rho \cdot d\tau +\sigma _0\text{
,\thinspace \thinspace \thinspace \thinspace \thinspace \thinspace
\thinspace \thinspace \thinspace \thinspace \thinspace \thinspace \thinspace
\thinspace \thinspace \thinspace \thinspace \thinspace \thinspace \thinspace
\thinspace \thinspace \thinspace \thinspace \thinspace }\rho _0=\text{ const.%
}  \label{3.32}
\end{equation}

\subsubsection{Euler-Lagrange's equations for the metric tensor field $g$}

The pressure $p$ is also a degenerated Lagrangian invariant with respect to
the field variables $g_{ij}$. The corresponding Euler-Lagrange equations for 
$g_{ij}$ could be written in the form 
\begin{equation}
\frac{\partial p}{\partial g_{ij}}+\frac 12\cdot p\cdot g^{\overline{i}%
\overline{j}}=0\text{ ,}  \label{3.33}
\end{equation}

\noindent where 
\begin{equation}
\frac{\partial p}{\partial g_{ij}}=\rho \cdot \frac{\partial f}{\partial
g_{ij}}+(u\rho )\cdot \frac{\partial b}{\partial g_{ij}}\text{ ,}
\label{3.34}
\end{equation}
\begin{equation}
\frac{\partial f}{\partial g_{ij}}=a_0\cdot u^{\overline{i}}\cdot u^{%
\overline{j}}+h_0\cdot (a^{\overline{i}}.\xi ^{\overline{j}}+a^{\overline{j}%
}.\xi ^{\overline{i}})\text{ ,\thinspace \thinspace \thinspace \thinspace
\thinspace \thinspace \thinspace \thinspace \thinspace \thinspace \thinspace
\thinspace \thinspace \thinspace \thinspace }\frac{\partial b}{\partial
g_{ij}}=\frac 12\cdot h_0\cdot (u^{\overline{i}}\cdot \xi ^{\overline{j}}+u^{%
\overline{j}}\cdot \xi ^{\overline{i}})\text{ .}  \label{3.35}
\end{equation}

The explicit form of the Euler-Lagrange equations for $g_{ij}$ follows in
the form 
\begin{eqnarray}
g^{ij} &=&-\frac 2p\cdot [a_0\cdot \rho \cdot u^i\cdot u^j+\frac 12\cdot
h_0\cdot \rho \cdot (a^i\cdot \xi ^j+a^j\cdot \xi ^i)+  \notag \\
&&+\frac 12\cdot h_0\cdot (\rho _{,m}\cdot u^m)\cdot (u^i\cdot \xi
^j+u^j\cdot \xi ^i)]\text{ .}  \label{3.36}
\end{eqnarray}

Now, we can write down in a table the Euler-Lagrange equations for the field
variables $\rho $, $u$, $\xi $, and $g$

\begin{center}
\begin{tabular}{cc}
Field variable & Euler-Lagrange's equations \\ 
\_\_\_\_ & \_\_\_\_\_\_\_\_\_\_\_\_\_\_\_\_\_\_\_\_\_\_\_\_\_\_ \\ 
$\rho $ & $l_{,i}\cdot u^{i}=g_{\overline{k}\overline{l}}\cdot \lbrack \frac{%
a_{0}}{h_{0}}\cdot u^{k}\cdot u^{l}+a^{k}\cdot \xi ^{l}+(q_{m}\cdot
u^{m}-u^{m}\,_{;m})\cdot u^{k}\cdot \xi ^{l}]$ \\ 
&  \\ 
$u$ & $\xi ^{i}\,_{;j}\cdot u^{j}=l\cdot (log\,\rho )_{,j}\cdot
g^{ji}+2\cdot \frac{a_{0}}{h_{0}}\cdot u^{i}+(q-\delta u)\cdot \xi ^{i}+g_{%
\overline{k}\overline{l}}\cdot u^{k}\,_{;j}\cdot g^{ji}\cdot \xi ^{l}-$ \\ 
& $-g^{ij}\cdot (g_{\overline{j}\overline{k}})_{;m}\cdot u^{m}\cdot \xi
^{k}+g_{k;j}^{i}\cdot u^{j}\cdot \xi ^{k}\text{ }$ \\ 
&  \\ 
$\xi $ & $a^{i}=u^{i}\,_{;j}\cdot u^{j}=-[(log\,\rho )_{,j}\cdot u^{j}]\cdot
u^{i}=-\frac{1}{\rho }\cdot (\rho _{,m}\cdot u^{m})\cdot u^{i}$ \\ 
&  \\ 
$g$ & $g^{ij}=-\frac{2}{p}\cdot \lbrack a_{0}\cdot \rho \cdot u^{i}\cdot
u^{j}+\frac{1}{2}\cdot h_{0}\cdot \rho \cdot (a^{i}\cdot \xi ^{j}+a^{j}\cdot
\xi ^{i})+$ \\ 
& $+\frac{1}{2}\cdot h_{0}\cdot (\rho _{,m}\cdot u^{m})\cdot (u^{i}\cdot \xi
^{j}+u^{j}\cdot \xi ^{i})]\text{ }$%
\end{tabular}
\end{center}

\subsection{Consequences from the Euler-Lagrange equations}

Let us now consider the system of the Euler-Lagrange equations for $\rho $, $%
u$, $\xi $, and $g$.

From (\ref{3.29}) and (\ref{3.36}), it follows that 
\begin{equation}
g^{ij}=-\frac{2\cdot \rho }p\cdot a_0\cdot u^i\cdot u^j\text{ .}
\label{3.37}
\end{equation}

After contraction of $g^{ij}$ with $u_{\overline{i}}\cdot u_{\overline{j}}$,
we have 
\begin{equation}
e=-\frac{2\cdot \rho }p\cdot a_0\cdot e^2:\,\,\,\,\,p=-2\cdot \rho \cdot
a_0\cdot e\text{ .}  \label{3.38}
\end{equation}

After contraction of $g^{ij}$ with $g_{\overline{i}\overline{j}}$, it
follows 
\begin{equation}
g^{ij}\cdot g_{\overline{i}\overline{j}}=n=-\frac{2\cdot \rho }p\cdot
a_0\cdot e:\,\,\,\,n\cdot p=-2\cdot \rho \cdot a_0\cdot e\text{ .}
\label{3.39}
\end{equation}

From the last two expressions for $p$, we obtain 
\begin{equation}
(n-1)\cdot p=0\text{ .}  \label{3.40}
\end{equation}

Therefore, if the Euler-Lagrange equations for $g_{ij}$ have to be
fulfilled, then either the dimension $dimM$ of the manifold $M$ should be
equal to 1 $(dim\,M=1)$ or the pressure $p$ should vanish. The first case $%
(n=1)$ leads to consideration of an auto-parallel curve as one-dimensional
manifold and a tangent to it vector $u$. In the second case $(p=0)$, we
obtain the relation $\rho \cdot a_0\cdot e=0$ with $\rho \neq 0$, $e\neq 0$.
Then only $a_0$ should vanish $(a_0=0)$, and $g^{ij}$ (respectively $g_{ij}$%
) could not be determined uniquely.

Therefore, if we wish to investigate a Lagrangian system with the given
Lagrangian invariant (\ref{3.11}) in a manifold $M$ with $dim\,M>1$, with $%
p\neq 0$, and determined metric tensor $g$, we should consider $g$ as a
given non-dynamic field variable or we can choose one of the following
possibilities:

(a) We should consider some of the other field variables $(u$, $\xi $, $\rho
)$ as non-dynamic (fixed, given) field variables.

(b) We should add additional terms to $p$ leading to well determined
Euler-Lagrange equations for $g$ (as it is the case in the Einstein theory
of gravitation, where $p_E:=p+c_0\cdot R$, $c_0=$ const., $R=g^{ij}\cdot
R_{ij}=g^{ij}\cdot g_l^k\cdot R^l{}_{ijk}$).

At the same time, from (\ref{3.18}) and (\ref{3.29}), the condition for $%
l=g(u,\xi )$ in the case $a_0=0$ follows in the form 
\begin{equation}
ul+h_0\cdot [u(log\,\rho )+(\delta u-q)]\cdot l=\frac{a_0}{h_0}\cdot e=0%
\text{ ,}  \label{3.41}
\end{equation}

\noindent allowing the trivial solution $l=0$, i.e. $u$ and $\xi $ could be
orthogonal to each other if $p=0$.

\subsubsection{Euler-Lagrange's equations for $\protect\rho $, $u$, and $%
\protect\xi $}

If we consider only the field variables $\rho $, $u$, and $\xi $ as dynamic
field variables, then the corresponding Euler-Lagrange equations (ELEs) will
have the forms

\begin{tabular}{cc}
Field variable & Euler-Lagrange's equations \\ 
\_\_\_\_ & \_\_\_\_\_\_\_\_\_\_\_\_\_\_\_\_\_\_\_\_\_\_\_\_\_\_\_\_\_\_ \\ 
$\rho $ & $ul+[u(log\,\rho )+(\delta u-q)]\cdot l=\frac{a_{0}}{h_{0}}\cdot e$
\\ 
&  \\ 
$u$ & $\xi ^{i}\,_{;j}\cdot u^{j}=l\cdot (log\,\rho )_{,j}\cdot
g^{ji}+2\cdot \frac{a_{0}}{h_{0}}\cdot u^{i}+(q-\delta u)\cdot \xi ^{i}+g_{%
\overline{k}\overline{l}}\cdot u^{k}\,_{;j}\cdot g^{ji}\cdot \xi ^{l}-$ \\ 
& $-g^{ij}\cdot (g_{\overline{j}\overline{k}})_{;m}\cdot u^{m}\cdot \xi
^{k}+g_{k;j}^{i}\cdot u^{j}\cdot \xi ^{k}\text{ }$ \\ 
&  \\ 
$\xi $ & $a^{i}=u^{i}\,_{;j}\cdot u^{j}=-[(log\,\rho )_{,j}\cdot u^{j}]\cdot
u^{i}=-\frac{1}{\rho }\cdot (\rho _{,m}\cdot u^{m})\cdot u^{i}$%
\end{tabular}

Since the Euler-Lagrange equations are valid for every preliminary given
affine connections and metrics, we can search for their solutions in
different type of spaces with affine connections and metrics. If $u$ is a
tangent vector $u=\frac d{d\tau }$ of a curve with parameter $\tau $, then
the Euler-Lagrange equation for $\rho $ appears as an ordinary differential
equation of first order for the scalar product $l=g(u,\xi )$ of $u$ and $\xi 
$%
\begin{equation}
\frac{dl}{d\tau }+g(\tau )\cdot l=\overline{k}_0\cdot e\text{ ,}
\label{3.42}
\end{equation}

\noindent where 
\begin{equation}
g(\tau )=[u(log\,\rho )+(\delta u-q)]\text{ ,\thinspace \thinspace
\thinspace \thinspace \thinspace \thinspace \thinspace \thinspace \thinspace
\thinspace \thinspace \thinspace \thinspace \thinspace \thinspace \thinspace 
}\overline{\text{\thinspace }k}_0=\frac{a_0}{h_0}=\text{const.}\neq 0\text{ .%
}  \label{0.35}
\end{equation}

The exact solution of the equation (\ref{3.42}) for $l(\tau )$ is : 
\begin{equation}
l\left( \tau \right) =exp(-\frac 12\cdot g\cdot \tau ^2)\cdot [\overline{k}%
_0\cdot e\cdot \int exp(\frac 12\cdot g\cdot \tau ^2)\,\cdot d\tau +c_1]%
\text{ , \thinspace \thinspace \thinspace \thinspace \thinspace \thinspace
\thinspace \thinspace \thinspace \thinspace \thinspace \thinspace \thinspace
\thinspace \thinspace \thinspace }c_1=\text{ const.}  \label{3.43}
\end{equation}

\textit{Special case:} $a_0:=0:p=p_0+h_0\cdot g(\nabla _u\rho u,\xi
)=p_0+h_0\cdot [(u\rho )\cdot l+\rho \cdot g(a,\xi )]$.

The Euler-Lagrange equations for $\rho $, $u$, and $\xi $ will have the form 
\begin{equation}
ul+[u(log\,\rho )+(\delta u-q)]\cdot l=0\text{ ,}  \label{3.44}
\end{equation}
\begin{equation}
a=\nabla _uu=-[u(log\,\rho )]\cdot u\text{ \thinspace ,}  \label{3.45}
\end{equation}
\begin{equation}
\nabla _u\xi =l\cdot \overline{g}(log\,\rho )+(q-\delta u)\cdot \xi +(\xi
)(g)(k_0)-\overline{N}  \label{3.46}
\end{equation}

\noindent where 
\begin{equation}
k_0=u^i\,_{;l}\cdot g^{lj}\cdot \partial _i\otimes \partial _j\text{
\thinspace \thinspace \thinspace \thinspace \thinspace \thinspace \thinspace
\thinspace \thinspace , \thinspace \thinspace \thinspace \thinspace
\thinspace \thinspace \thinspace \thinspace \thinspace \thinspace \thinspace
\thinspace \thinspace \thinspace }\overline{N}=[g^{ij}\cdot (g_{\overline{j}%
\overline{k}})_{;m}\cdot u^m-g_{k;j}^i\cdot u^j]\cdot \xi ^k\cdot \partial _i%
\text{ .}  \label{3.47}
\end{equation}

If the Euler-Lagrange equations (\ref{3.44}) $\div \,$(\ref{3.46}) are
fulfilled, then $p=p_0=$ const. The auto-parallel equation for $u$ appears
in a non-canonical form. The equation (\ref{3.44}) has as trivial solution $%
l=0$, i.e. $\xi $ could be chosen as an orthogonal to $u$ contravariant
vector field.

\textit{Special case}: $\rho =\rho _0:=$ const.$\,\neq 0:p=p_0+a_0\cdot \rho
_0\cdot e+h_0\cdot \rho _0\cdot g(a,\xi )$.

The Euler-Lagrange equations for $u$ and $\xi $ will have the form 
\begin{equation}
\nabla _uu=a=0\text{ ,}  \label{3.48}
\end{equation}
\begin{equation}
\nabla _u\xi =2\cdot \frac{a_0}{h_0}\cdot u+(q-\delta u)\cdot \xi +(\xi
)(g)(k_0)-\overline{N}\text{ \thinspace \thinspace .}  \label{3.49}
\end{equation}

If (\ref{3.48}) and (\ref{3.49}) are fulfilled, then $p=p_0+a_0\cdot \rho
_0\cdot e$. The auto-parallel equation appears in its canonical form. If $u$
is a normalized vector field [$g(u,u)=e=$ const.], then $p=$ const.
Therefore, a Lagrangian systems of particles with constant rest mass density 
$\rho _0$ could exists moving on auto-parallel trajectories in a space with
pressure $p$ proportional to the kinetic energy of the particles. The
contravariant vector field $\xi $ is transported along $u$ in a special way
depending on the characteristics of $u$ and the $(\overline{L}_n,g)$-space.

\textit{Special case}: $\rho =\rho _0:=$ const.$\,\neq 0$, $%
a_0=0:p=p=p_0+h_0\cdot \rho _0\cdot g(a,\xi )$.

The Euler-Lagrange equations for $u$ and $\xi $ will have the form 
\begin{equation}
\nabla _uu=a=0\text{ ,}  \label{0.36}
\end{equation}
\begin{equation}
\nabla _u\xi =(q-\delta u)\cdot \xi +(\xi )(g)(k_0)-\overline{N}\text{
\thinspace \thinspace .}  \label{3.50}
\end{equation}

If the Euler-Lagrange equations are fulfilled, then $p=p_0=$ const. The
auto-parallel equation appears in its canonical form. Therefore, a
Lagrangian system of particles with constant rest mass density $\rho _0$
could exist moving on auto-parallel trajectories in a space with constant
pressure $p$.

\subsubsection{Auto-parallel equation as Euler-Lagrange's equation related
to a frame of reference}

If we use the basic arguments for introducing a generalized definition of a
frame of reference $(FR)$ [The set $FR\sim [u$, $T^{\perp u}(M)$, $\nabla
=\Gamma $, $\nabla _u]$ is called frame of reference \cite{Manoff-7} in a
differentiable manifold $M$ considered as a model of the space or of the
space-time] we can also find a solution of the G-A problem by the use of the
method of Lagrangians with covariant derivatives (MLCD).

Let us define a Lagrangian invariant in the form 
\begin{eqnarray}
L &=&p_0+h_0\cdot g[\nabla _u(\rho \cdot u)\text{,}\xi ]=p_0+h_0\cdot g_{%
\overline{i}\overline{j}}\cdot (\rho u^i)_{;k}\cdot u^k\cdot \xi ^j\text{ ,}
\notag \\
\text{ \thinspace \thinspace \thinspace \thinspace }p_0\text{, \thinspace
\thinspace \thinspace \thinspace }h_0 &=&\text{ const., \thinspace
\thinspace }\rho \in C^r(M)\text{ , \thinspace \thinspace \thinspace
\thinspace \thinspace }u\text{, \thinspace }\xi \in T(M)\text{ .}\,\,
\label{3.2}
\end{eqnarray}

\noindent with the additional condition for the contravariant vector fields $%
u$ and $\xi $: $g(u,\xi )=l=0$. The corresponding action $S$ could be
written in the form 
\begin{equation}
S=\int \sqrt{-d_g}\cdot (L+\lambda \cdot l)\cdot d^{(n)}x=\int (L+\lambda
\cdot l)\cdot d\overline{\omega }\text{ , \thinspace \thinspace \thinspace
\thinspace \thinspace \thinspace \thinspace \thinspace }d_g=\det (g_{ij})<0\,%
\text{\thinspace \thinspace .}  \label{3.3}
\end{equation}

\noindent where $\lambda $ is a Lagrangian multiplier. $L$ is interpreted as
the pressure $p$ of a physical system, $u$ is the velocity of the particles, 
$\rho $ is their proper mass density, and $\xi $ is a vector, orthogonal to $%
u$. By the use of the MLCD we obtain the covariant Euler-Lagrange equations
for the vector fields $u$ and $\xi $ obeying the condition $l=0$%
\begin{equation}
\frac{\delta L}{\delta \xi ^i}=0:\,\,\,\,\,\,\,\,\,\,\,\,\,\,u^i\,_{;j}\cdot
u^j=[\frac \lambda {h_0}-u(\log \rho )]\cdot u^i\text{ ,}  \label{3.4}
\end{equation}
\begin{eqnarray}
\frac{\delta L}{\delta u^i} &=&0:\,\,\,\,\,\xi ^i\,_{;j}\cdot u^j=(q-\delta
u+\frac \lambda {h_0})\cdot \xi ^i+g_{\overline{k}\overline{l}}\cdot
u^k\,_{;j}\cdot g^{ji}\cdot \xi ^l-  \notag \\
&&-[g^{ij}\cdot (g_{\overline{j}\overline{k}})_{;m}\cdot u^m-g_{k;j}^i\cdot
u^j]\cdot \xi ^k\text{ .}  \label{3.5}
\end{eqnarray}
\begin{equation}
\frac{\delta L}{\delta \lambda }=0:g(u,\xi )=l=0\text{ .}  \label{3.5a}
\end{equation}

In index-free form the equations for $u$ and $\xi $ would have the forms: 
\begin{equation}
\nabla _uu=k\cdot u\text{ ,\thinspace \thinspace \thinspace \thinspace
\thinspace \thinspace \thinspace \thinspace \thinspace \thinspace \thinspace
\thinspace \thinspace \thinspace \thinspace \thinspace \thinspace \thinspace
\thinspace \thinspace }k=\frac \lambda {h_0}-u(\log \rho )\text{ ,}
\label{3.6}
\end{equation}
\begin{equation}
\nabla _u\xi =m\cdot \xi +K-\overline{N}\text{ , \thinspace \thinspace
\thinspace \thinspace \thinspace \thinspace \thinspace \thinspace \thinspace
\thinspace }m=q-\delta u+\frac \lambda {h_0}\text{ , \thinspace \thinspace
\thinspace \thinspace \thinspace \thinspace \thinspace \thinspace \thinspace 
}  \label{3.7}
\end{equation}
\begin{equation}
q=q_j\cdot u^j\text{ , \thinspace \thinspace \thinspace \thinspace }%
q_j=T_{kj}\,^k-\frac 12\cdot g^{\overline{k}\overline{l}}\cdot
g_{kl;j}+g_k^l\cdot g_{l;j}^k\text{ ,}  \label{3.8}
\end{equation}
\begin{equation}
\delta u=u^k\,_{;k}\text{ , \thinspace \thinspace \thinspace \thinspace
\thinspace }K=K^i\cdot \partial _i=(g_{\overline{k}\overline{l}}\cdot
u^k\,_{;j}\cdot g^{ji}\cdot \xi ^l)\cdot \partial _i\text{ ,}  \label{3.9}
\end{equation}
\begin{equation}
\overline{N}=\overline{N}\,^i\cdot \partial _i\text{ ,\thinspace \thinspace
\thinspace \thinspace \thinspace \thinspace }\overline{N}\,^i=[g^{ij}\cdot
(g_{\overline{j}\overline{k}})_{;m}\cdot u^m-g_{k;j}^i\cdot u^j]\cdot \xi ^k%
\text{ .}  \label{3.10}
\end{equation}

The Euler-Lagrange's equation (\ref{3.6}) is just the auto-parallel equation
in a non-canonical form. For $\rho =$ const., it will have the form $\nabla
_uu=\frac \lambda {h_0}\cdot u$. After changing the parameter of the curve
to which $u$ is a tangent vector field the auto-parallel equation could be
found in its canonical form $\nabla _{\overline{u}}\overline{u}=0$.

The Euler-Lagrange's equation for $\xi $ (\ref{3.7}) has in general a more
complicated form than the parallel equation for $\xi $ along $u$ ($\nabla
_u\xi =g\cdot \xi $). For different affine connections (and the
corresponding models of space-time) this equation would have different
solutions. Therefore, if we consider an auto-parallel equation as a result
of a variational principle we should take into account the corresponding
orthogonal to $u$ sub space.

\begin{remark}
The Lagrangian invariant $L$ could be defined without the requirement $\xi $
to be orthogonal to $u$. The covariant Euler-Lagrange's equations will be
then found for $u$ and a vector field $\xi \in T(M)$. For $\rho =$ const.
the auto-parallel equation will have its canonical form. The orthogonality
condition for $u$ and $\xi $ could be introduced after solving the
Euler-Lagrange equations for $u$ or $\xi $.
\end{remark}

\section{Conclusion}

In this paper we have considered the finding out of the auto-parallel
equation in spaces with affine connections and metrics by the use of
Lagrangian formalism related to the method of Lagrangians with covariant
derivatives. If an appropriate Lagrangian density is taken into account,
then the auto-parallel equation could be found as Euler-Lagrange's equation
for an auto-parallel transported contravariant vector field. It was already
shown that this type of equation could be interpreted as an equation for the
motion of a free moving test particles in a space with affine connections
(considered as a model of the space-time). The derivation of the
auto-parallel equation on the basis of a Lagrangian formalism proves once
more that this equation could be of use in field theories over more general
spaces than the (pseudo) Riemannian spaces with or without torsion.

\end{document}